# Modeling Growth Process of $\beta$ index of Transport Network Based on Nonlinear Spatial Dynamics

Yanguang Chen

(Department of Geography, College of Urban and Environmental Sciences, Peking University, 100871, Beijing, China. Email: chenyg@pku.edu.cn)

**Abstract:** The $\beta$ index is one of important measurements reflecting development level of traffic networks. However, how to explain and predict the $\beta$ index growth for a geographical region is a pending problem. With the help of mathematical reasoning and empirical analysis, this paper is devoted to modeling the growing curve of $\beta$ index. A new measurement termed $\delta$ index is introduced and a set of new models are constructed. Suppose there is a nonlinear relation between human settlements and roads. A pair differential equations are built for describing the nonlinear dynamics of traffic networks. A logistic function of $\beta$ index growth is derived from the two spatial dynamic equations. On the other hand, based on verified empirical models about urbanization level, economic development level, and $\beta$ index of traffic network, a Boltzmann equation of $\beta$ index growth can be derived. Normalizing the $\beta$ index, Boltzmann equation will become logistic function. This lends an indirect support to the theoretical model of $\beta$ index from the prospective of positive studies. The models proposed in this work provided new approaches for explanation and prediction of spatio-temporal evolution of traffic networks.

## 1 Introduction

A system of human settlements in a geographical area is organized by transportation and communication networks. Among these networks, the transportation network is tangible and visible, while the communication network is partially visible. Therefore, an urban system is defined as a network of cities and towns as well as their hinterlands (Mayhew, 1997). An urban system and a transportation network represents two different sides of the same coin for a regional system. The attributes and dynamics of urban systems reflect the most important aspects of urbanization (Knox and Marston, 2016). On the other hand, changes of traffic networks are correlated with urbanization process, and the relation can be mirrored by a simple linear function of traffic beta index and urbanization level (Chen, 2004). Urbanization conforms to some mathematical laws such as logistic growth and nonlinear dynamics (Cadwallader, 1996; Chen, 2009; Karmeshu, 1988; Pacione, 2009; Rao *et al*, 1989; United Nations, 1980; Zhou, 1995). Now we are concerned about whether the laws of urbanization, especially the mathematical rules of urban system, are reflected in transportation networks.

Scientific research consists of two processes: describing (*how* a systems work) and understanding (*why* the system work in this way). To describe spatial characteristics of traffic systems, we need a set of basic measurements, which are termed "indexes". To understand the mechanism of traffic evolution, we need a series of mathematical models based on the measurements. Among various traffic network indexes, the $\beta$ index is simple and easily worked out (Zeng, 1996; Zeng, 2001). Based on the concise index, this paper is devoted to modeling spatial dynamics of traffic networks. Because it is difficult to directly deduce the prediction model of $\beta$ index, we need some mathematical transformation approach. First, a new measurement termed $\delta$ index of traffic network is defined, being actually mathematical equivalent to $\beta$ index. Second, a system of dynamic equations are constructed for traffic networks based on the $\delta$ index. Third, a prediction function of the $\delta$ index is derived from the dynamic equation systems. The prediction model of $\delta$ index can be turned into a model for the $\beta$ index. The rest parts of this paper are organized as follows. In Section 2, a new index is proposed and new mathematical models are built and derived. In Section 3, an empirical chain of evidence is presented for the theoretical models. In Section 4, related questions are discussed, and finally, the discussion will be concluded by summarizing the main points of this study.

# 2 Modeling results

## 2.1 Basic measurements

A network comprises an ordered set of nodes and links, which is often termed as vertexes and edges in network theory. The essential elements of geographical systems can be abstracted as points, lines, and areas. For a network of human settlements in a geographical region (area), nodes include cities and towns (points), and links include roads, railways, waterways and airways (lines). Now, let's investigate a regional system from the perspective of transportation. To derive the prediction model of $\beta$ index of traffic network, we need a simple measurement, which can be defined by numbers of nodes and links. Based on temporal process, the $\beta$ index of traffic network is as follows

$$\beta(t)=\frac{c(t)}{v(t)}, \tag{1}$$

where $\beta(t)$ denotes $\beta$ index of time $t$, $c(t)$ refers to the number of links (roads), and $v(t)$ to the number of points (nodes) at time $t$. The $\beta$ index can be regarded as measurement of ratio of part to part in a traffic network. Correspondingly, define a new index as below:

$$\delta(t)=\frac{c(t)}{c(t)+v(t)}, \tag{2}$$

where $\delta(t)$ can be termed $\delta$ index of time $t$. The $\delta$ index can be treated as a measurement of proportion of part to whole in a traffic network. Thus we have

$$1-\delta(t)=\frac{v(t)}{c(t)+v(t)}. \tag{3}$$

The relation between the $\beta$ index and $\delta$ index is as follows

$$\frac{1}{\delta(t)}=\frac{c(t)+v(t)}{c(t)}=1+\frac{1}{\beta(t)}, \tag{4}$$

which is a hyperbolic function. This is a nonlinear relationship: $\delta=\beta/(\beta+1)$, $\beta=\delta/(1-\delta)$. This means that the two measures, $\beta$ index and $\delta$ index, are mathematically equivalent, but not numerically equivalent. For the initial value and capacity value, we have

$$\frac{1}{\delta_0}-1=\frac{1}{\beta_0}, \tag{5}$$

$$\frac{1}{\delta_{\max}}-1=\frac{1}{\beta_{\max}}. \tag{6}$$

These two equations will be used in subsequent mathematical reasoning for the models of transport network development.

## 2.2 Spatial dynamic model

Suppose there is a geographical region in which a traffic network consisting of towns and roads develops. There are relationships of competition and synergy between towns (nodes) and roads (links). On the one hand, towns and roads depend on one another (cooperative effect). On the other, roads and towns compete for resources and geographical space for their own development (competitive effect). Thus, the dynamical process of a traffic network can be expressed as a pair of differential equations as follows

$$\begin{cases} \frac{\mathrm{d}c(t)}{\mathrm{d}t} = Ac(t) + B\frac{c(t)v(t)}{c(t)+v(t)} \\ \frac{\mathrm{d}v(t)}{\mathrm{d}t} = Cv(t) - D\frac{c(t)v(t)}{c(t)+v(t)} \end{cases}, \tag{7}$$

where $A$, $B$, $C$, and $D$ denote parameters. In equation (7), the cross term, $c(t)v(t)/(c(t)+v(t))$, represents the collaboration or coupling relationship. If the coupling term in equation (7) is removed, the rest parts becomes the allometric growth equations, which represents a spatial competitive relationship.

Based on the two measurements of traffic connectivity and the relationship between them, a logistic model can be derived from the spatial dynamic model of traffic network. Differentiating equation (2) with respect to time $t$ yields

$$\frac{\mathrm{d}\delta(t)}{\mathrm{d}t} = \frac{\mathrm{d}c(t)/\mathrm{d}t}{c(t)+v(t)} - \frac{c(t)}{[c(t)+v(t)]^2}\left[\frac{\mathrm{d}c(t)}{\mathrm{d}t} + \frac{\mathrm{d}v(t)}{\mathrm{d}t}\right]. \tag{8}$$

Substituting equation (7) into equation (8) yields

$$\frac{\mathrm{d}\delta(t)}{\mathrm{d}t} = \frac{Ac(t) + B\frac{c(t)v(t)}{c(t)+v(t)}}{c(t)+v(t)} - \frac{c(t)}{[c(t)+v(t)]^2}\left[Ac(t) + Cv(t) + (B-D)\frac{c(t)v(t)}{c(t)+v(t)}\right]. \tag{9}$$

Reorganizing equation (9), we have

$$\frac{\mathrm{d}\delta(t)}{\mathrm{d}t} = \frac{Ac(t)}{c(t)+v(t)} - \frac{Ac(t)^2}{[c(t)+v(t)]^2} + (B-C)\frac{c(t)v(t)}{[c(t)+v(t)]^2}. \tag{10}$$

Substituting equations (2) and (3) into equation (10) produces a differential equation as follows

$$\frac{\mathrm{d}\delta(t)}{\mathrm{d}t} = (A+B-C)\delta(t)(1-\delta(t)). \tag{11}$$

The solution of equation (10) is a logistic function, that is

$$\delta(t)=\frac{1}{1+(1/\delta_0-1)e^{-kt}}, \tag{12}$$

in which the parameter $k$ can be expressed as

$$k=A+B-C. \tag{13}$$

In equation (12), $k$ denotes the inherent growth rate of the $\delta$ value of transport network.

## 2.3 Derivation of logistic model of *β* index

The common measurement of traffic connectivity is the well-known $\beta$ index. It can be easily proved that if the $\delta$ index follow a logistic relation, the $\beta$ index also follow a logistic relation. Substituting equation (12) into equation (4) yields and exponential growth equation.

$$\beta(t)=\frac{\delta(t)}{1-\delta(t)}=\beta_0 e^{-kt}. \tag{14}$$

where $\beta_0=\delta_0/(1-\delta_0)$, which can be derived from equation (5). This indicates that if the $\delta$ index curve satisfy a two-parameter logistic function, the $\beta$ index curve will satisfy an exponential growth function. If so, it suggests that $v(t)\rightarrow 0$, or $c(t)\rightarrow\infty$. However, in practice, this is impossible. To describe the developing process of transport network, the two-parameter logistic function, equation (12), can be revised as a three-parameter logistic function as below

$$\delta(t)=\frac{\delta_{\max}}{1+(\delta_{\max}/\delta_0-1)e^{-kt}}. \tag{15}$$

Substituting equation (15) into equation (4) yields

$$\beta(t)=\frac{1}{1/\delta_{\max}-1+(1/\delta_0-1/\delta_{\max})e^{-kt}}. \tag{16}$$

Substituting equations (5) and (6) into equation (16) yields

$$\beta(t)=\frac{1}{1/\beta_{\max}+(1/\beta_0-1/\beta_{\max})e^{-kt}}=\frac{\beta_{\max}}{1+(\beta_{\max}/\beta_0-1)e^{-kt}}. \tag{17}$$

This suggests, if the $\delta$ index curve satisfies a three-parameter logistic function, the $\beta$ index curve will also satisfy a three-parameter logistic function. Equation (7) can used to explain the evolution of traffic networks, equation (15) and (17) can utilize to prediction the growth of traffic networks. Further, replacing the time dummy variable with a set of real explanatory variables $x_j$, that is, let

$kt=\sum b_j x_j$, we have

$$\beta(t)=\frac{\beta_{\max}}{1+(\beta_{\max}/\beta_0-1)\exp(-\sum_{j=1}^{n}b_j x_j)},\tag{18}$$

where $j$=1, 2,…, $m$, $b_j$ denotes parameters, $m$ represents the number of explanatory variables. Using equation (18), maybe we can develop an approach of logistic regression analysis for revealing the influence factors on evolution of traffic networks.

# 3 Empirical evidences

## 3.1 Basic evidences

In order to directly verify the above models, we need to have the time series of $\beta$ indexes of traffic connectivity based on observation data (nodes and roads). However, no observation data for many consecutive years has been obtained. In this case, the empirical models established by predecessors can be used to indirectly verify the above theoretical derivation results. The first model is about the mathematical relationship between per capita income and $\beta$ indexes of traffic network, which can be described with a logarithmic function as follows (Taylor, 1977)

$$\beta=\kappa\ln x+\phi,\tag{19}$$

where $\beta$ denotes the $\beta$ index of connectivity, $x$ is to per capita income for selected countries, and $\kappa$, $\varphi$ are two parameters. Equation (19) was confirmed by Taylor (1977). For the relationships between railway connectivity and the per capita income of selected countries, the empirical model is as below:

$$\hat{\beta}=0.197+0.381\log x=0.197+0.165\ln x.\tag{20}$$

The goodness of fit is about $R^2$=0.5112 (Taylor, 1977).

The second model is about the mathematical relationship between per capita income and level of urbanization. The relationship can also be characterized with a logarithmic function as below (Zhou, 1989)

$$L=a\ln x-b=a\ln(\mu z^{\alpha})-b=a'\ln z-b',\tag{21}$$

where $L$ denotes the level of urbanization, $x$ is per capita income, $z$ is per capita outcome, and $a$, $b$, $\alpha$, $\mu$, $a'=a\alpha$,$b'=a\ln\mu-b$ are six parameters. Equation (21) was verified by Zhou (1989) (Figure 1). In equations (19), (20), and (21), the per capita income can be replaced by per capita economic output such as per capita gross domestic product (GDP). The relationship between urbanization level and per capita gross national product (GNP) of 157 countries in 1977 are as follows

$$\hat{L} = -74.6799 + 40.4550\log x = -74.6799 + 17.5694\ln x\,. \tag{22}$$

The goodness of fit is about $R^2$=0.9227 (Zhou, 1995).

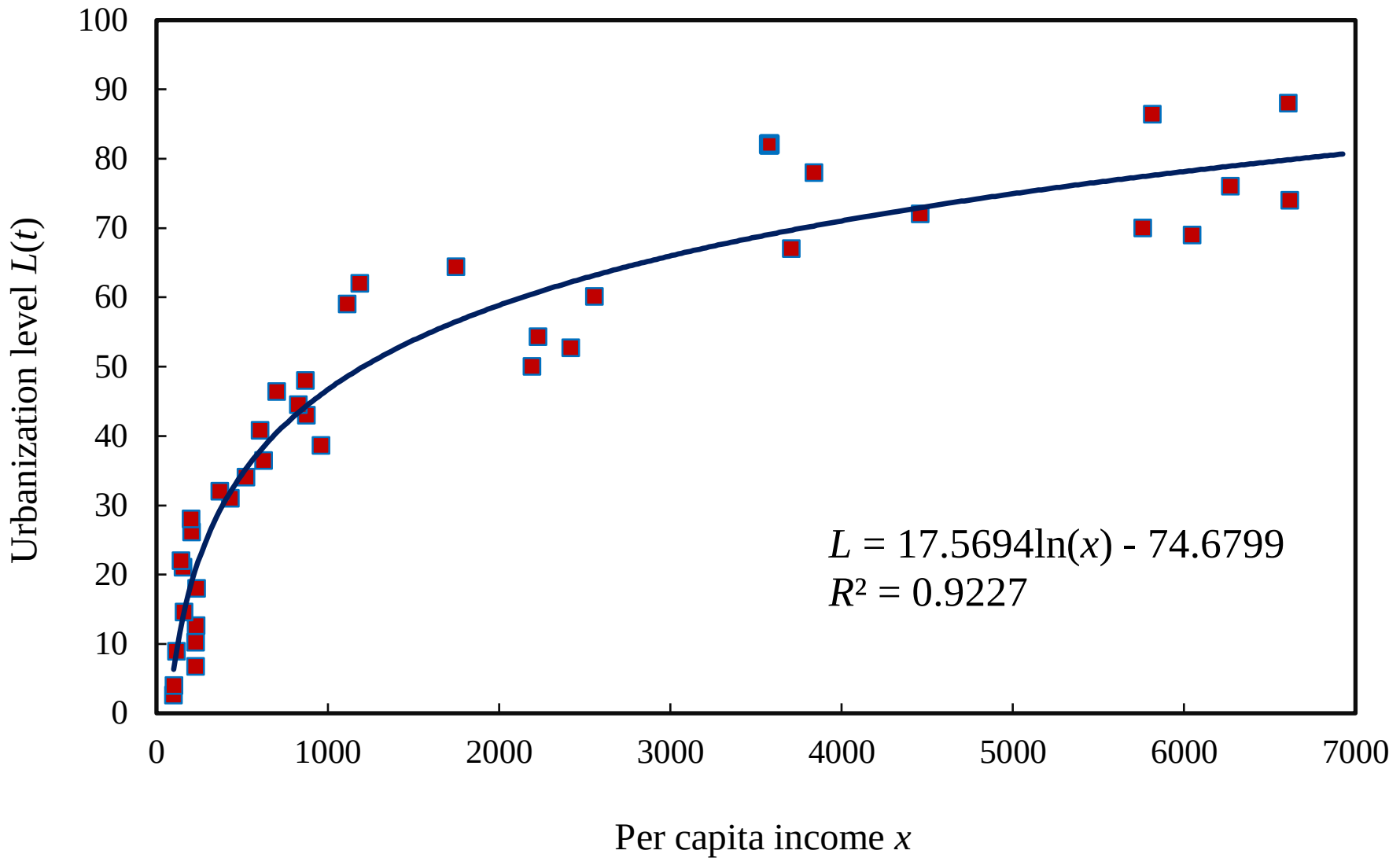

**Figure 1 The logarithmic relationship between urbanization level and economic development level based on per capita income the countries around the world (1977)**

(**Note**: The original data come from Professor Yixing Zhou, who processed the data of urbanization level and per capita GNP of 157 countries all over the world in 1977. The estimated values of parameters are subtly different from the results of Zhou (1995) due to recalculating.)

The third model of the logistic model of urbanization curve. An urbanization curve is the curve of urbanization level growth (Cadwallader, 1996; Pacione, 2009). The curve can be expressed as (United Nations, 1980; Zhou, 1995)

$$L(t) = \frac{L_{\max}}{1 + (L_{\max} / L_0 - 1)e^{-kt}}\,, \tag{23}$$

where $L(t)$ denotes the level of urbanization of a geographical region at time $t$ ($t=n-n_0=0,1,2,\ldots$, here $n$ refers to year, and $n_0$ to the start year), $L_0$ is the initial value of urbanization level at the start year, $L_{\max}$ is the capacity value of urbanization level, and $k$ refers to the inherent growth rate of urbanization level. Equation (23) can be verified by the observation data of the United States of America (Figure 2). Based on the urban and rural census data of USA from 1790 to 2010, the model is as below:

$$L(t) = \frac{86.6868}{1 + 19.6016e^{-0.0252t}}, \quad (24)$$

where $t=n$-1790 is time, and $n$ represents year. The goodness of fit is about $R^2$=0.9946.

If the per capita income is replaced by per capita economic output such as per capita GDP, the logarithmic relation of $\beta$ index or urbanization level to the economic variable will not change. In other words, the relation between per capita economic output and $\beta$ index or urbanization level still satisfies logarithmic function. The reason is that the relationships between per capita income and various per capita outputs obeys the allometric scaling law (Chen, 2010). That is, the empirical relation between per capita income and per capita economic output is as below

$$x = \mu z^{\alpha}, \quad (25)$$

where $x$ refers to per capita income, and $z$ to per capita economic output such as per capita GDP, $\mu$ is a proportionality coefficient, and $\alpha$ is a scaling exponent. According to the provincial observation data in China, we have $\mu$≈0.5, $\alpha$≈1.05 (Chen, 2010). For example, substituting equation (25) into equation (19) yields

$$\beta = \kappa \ln(\mu z^{\alpha}) + \phi = \alpha\kappa \ln(z) + \kappa \ln(\mu) + \phi = \kappa' \ln(z) + \phi', \quad (26)$$

where $\kappa'=\alpha\kappa$ and $\phi'=\kappa\ln(\mu)+\phi$ are two parameters. This indicates that using other per capita economic variables to take the place of per capita income does not change the logarithmic forms of equations (19) and (21), and thus does not change the logarithmic forms of equations (20) and (22).

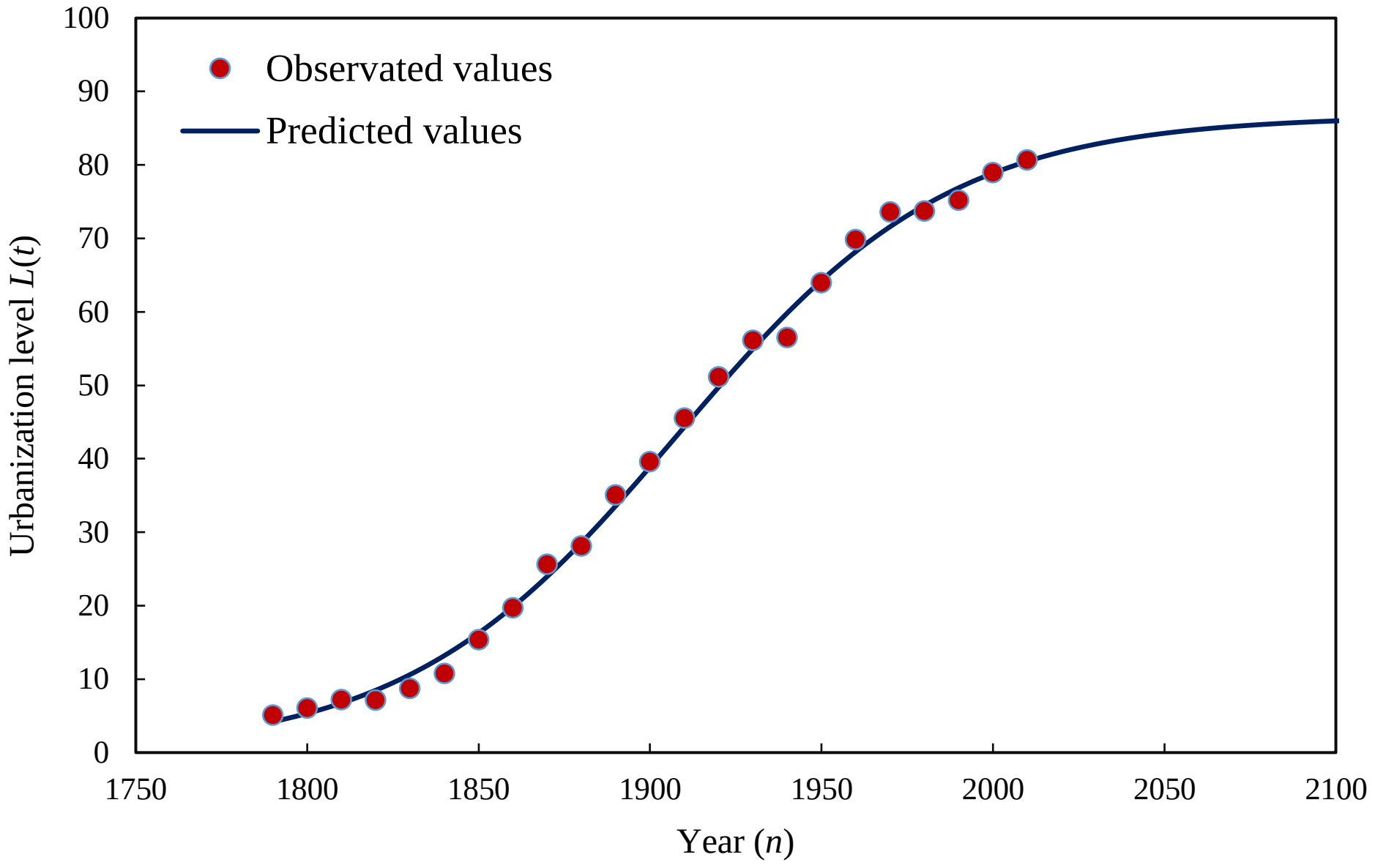

**Figure 2 The scattered points for observed values and the curve for predicted value of U.S. urbanization level (1790-2010)**

(**Note**: The level of urbanization of the United States of America can be modeled with a logistic function. Based on the census data from 1790 to 2010, the parameters are as follows: $k$=0.0252, $L_{max}$=86.6868%, $L_0$=4.2078%. The goodness of fit is about $R^2$=0.9946.)

## 3.2 Indirect proof

The logistic model of $\beta$ indexes of traffic connectivity can be derived from the three empirical models abovementioned. Combining equation (19) and (21) yields (Chen, 2004)

$$\beta = \kappa(\frac{L+b}{a}) + \phi = \omega L + \vartheta , \tag{27}$$

where $\omega=\kappa/a$ and $\vartheta=a\kappa b/a+\varphi$ are two parameters. Substituting equation (23) into equation (27) yields

$$\beta = \kappa(\frac{L+b}{a}) + \phi = \frac{\omega L_{\max}}{1+(L_{\max}/L_0-1)e^{-kt}} + \vartheta . \tag{28}$$

For simplicity and explicitness, let

$$\vartheta = \beta_{\min} , \quad \omega L_{\max} = \beta_{\max} - \beta_{\min} , \quad \omega L_0 = \beta_0 - \beta_{\min} . \tag{29}$$

Then equation (28) can be re-expressed as

$$\beta(t) = \beta_{\min} + \frac{\beta_{\max} - \beta_{\min}}{1+(\beta_{\max}-\beta_0)/(\beta_0-\beta_{\min}))e^{-kt}} , \tag{30}$$

where $\beta_{min}$ is the minimum value of the $\beta$ index, the other symbols are the same as those in equation (17). This is the mathematical form of Boltzmann equation, which can be used to model urban phenomena (Benguigui *et al*, 2001). If the $\beta_{min}$ value is very small, equation (30) can be simplified to the logistic function, equation (17). In fact, normalizing the $\beta$ index, we can transform equation (30) into a logistic function as follows

$$\beta^*(t) = \frac{\beta(t)-\beta_{\min}}{\beta_{\max}-\beta_{\min}} = \frac{1}{1+(1/\beta_0^*(t)-1)e^{-kt}} , \tag{31}$$

where $\beta^*(t)$ =($\beta(t)$-$\beta_{min}$)/($\beta_{max}$-$\beta_{min}$) denotes the normalized $\beta$ index at time $t$, the parameter indicating the initial value can be expressed as $\beta_0^*(t)$=($\beta_0$-$\beta_{min}$)/($\beta_{max}$-$\beta_{min}$). Anyway, the logistic function can be regarded as a special case or an approximate form of Boltzmann equation. Since the logistic model of the $\beta$ index can be derived from the empirical models which have been demonstrated, the empirical basis of the model is of no problem. Based on the empirical models, equation (20), equation (22), equation (24), and equation (25), an approximate empirical model can constructed as follows

$$\hat{\beta}=0.8208+\frac{0.8572}{1+19.6016e^{-0.0252t}}\approx\frac{1.6780}{1+19.6016e^{-0.0252t}}\ . \tag{32}$$

Of course, this model is for reference only. In light of equations (5), (6), and (15), an approximate logistic model for $\delta$ index can be derived from equation (32) as below:

$$\hat{\delta}=\frac{0.6266}{1+7.3194e^{-0.0252t}}\ . \tag{33}$$

If a reader has a time series data of traffic connectivity for a geographical region, he can estimate the model parameters in the better way.

In geography, a mathematical law based on time series is often consistent with that based on spatial series and hierarchical series. However, for transport network, thing seems to be different. Based on cross-sectional data, the relationships between urbanization level and the $\beta$ index of transport network can be described with a linear function, equation (27). However, this relation cannot be always generalized to the time series data. From equations (17) and (23) it follows

$$\frac{\beta(t)}{\beta_{\max}-\beta(t)}=\eta(\frac{L(t)}{L_{\max}-L(t)})^{\sigma}\ , \tag{34}$$

in which $\eta$ refers to the proportionality coefficient, and $\sigma$ to the scaling exponent. Equation (34) can be regarded as an allometric scaling relation based on logistic processes (Chen, 2014). The common allometric scaling is based on exponential growth (Bertalanffy, 1968). Two exponential laws compose a power law (Chen, 2015). However, for the logistic growth, the allometric scaling is latent and complex, and can be expressed by the mathematical form similar to equation (34). This suggests that, for transport networks, the spatio-temporal relationships are more complicated than urban systems. The above derivation and empirical analysis are based on a simple case, i.e., equation (27).

# 4 Questions and discussion

The theoretical derivation and empirical evidences lead to the following inferences on traffic network. First, there are a pair of basic measurements for traffic connectivity, that is, $\beta$ index and $\delta$ index. The two indexes can be linked by a hyperbolic function. Second, both the time series of $\beta$ index and that of $\delta$ index can be modeled with Boltzmann equation and logistic function. The logistic function can be treated as special case of Boltzmann equation. Third, the logistic models of $\beta$ index and $\delta$ index growth can be derived from the dynamic equations of spatial interaction between cities and roads. The dynamic model is a pair of differential equation describing growing process of systems of cities and roads. In short, the spatial dynamics can be employed to explain development of traffic network, and the logistic function can be utilized to describe and predict the growth of traffic connectivity in a geographical region.

The logistic models of connectivity can be used to define and research the growing stages of urban-traffic network. Taking the first derivative of equation (17) with respect to time $t$ yields a speed formula of the $\beta$ index as follows

$$S(t)=\frac{\mathrm{d}\beta(t)}{\mathrm{d}t}=k\beta(t)(1-\frac{\beta(t)}{\beta_{\max}}), \tag{35}$$

where $S(t)$ denotes the growth rate of the $\beta$ index at time $t$. Equation (35) can give the unimodal curve of growth rate of the $\beta$ index (Figure 3). For urban and traffic systems, a growth rate can be treated as a type of growing speed. Taking the first derivative of equation (35) with respect to time $t$, or taking the second derivative of equation (17), yields an acceleration formula of the traffic growth as below

$$a(t)=\frac{\mathrm{d}^2\beta(t)}{\mathrm{d}t^2}=\frac{\mathrm{d}S(t)}{\mathrm{d}t}=k^2\beta(t)(1-\frac{2\beta(t)}{\beta_{\max}})(1-\frac{\beta(t)}{\beta_{\max}}), \tag{36}$$

where $a(t)$ denotes the growing acceleration of the $\beta$ index at time $t$. Equation (36) can give the inverse S-shaped curve of growing acceleration of the $\beta$ index (Figure 4). From equation (35) we can derive a formula for the second inflexion point, i.e., the middle division point. From equation (36) we can derive two formulae for the first inflexion point and the third inflexion point, namely, the lower division point and the upper inflexion point, respectively. Thus, the growing process of a transport network can be divided into four stages based on the $\beta$ index: initial slow growth, accelerated fast growth, decelerated fast growth, and terminal slow growth (Table 1).

**Table 1 Phase division results and inflexion points of transport network development based on logistic model of the $\beta$ index**

| Scheme (I) | Scheme (II) | Scheme (III) | Middle limit | Lower and upper limits |
|---|---|---|---|---|
| Acceleration | Initial slow growth | Initial slow growth | | $\beta_1(t)=(\frac{1}{2}-\frac{1}{2\sqrt{3}})\beta_{\max}$ |
| | Celerity (fast | Accelerated | | |
| | growth) | fast growth | $\beta_{\mathrm{m}}(t)=\frac{\beta_{\max}}{2}$ | |
| Deceleration | | Decelerated fast growth | | $\beta_{\mathrm{u}}(t)=(\frac{1}{2}+\frac{1}{2\sqrt{3}})\beta_{\max}$ |
| | Terminal slow growth | Terminal slow growth | | |

**Note**: In the table, $\beta_{\mathrm{l}}(t)$ refers to the lower division, $\beta_{\mathrm{m}}(t)$ to the middle division, and $\beta_{\mathrm{u}}(t)$ to the upper division.

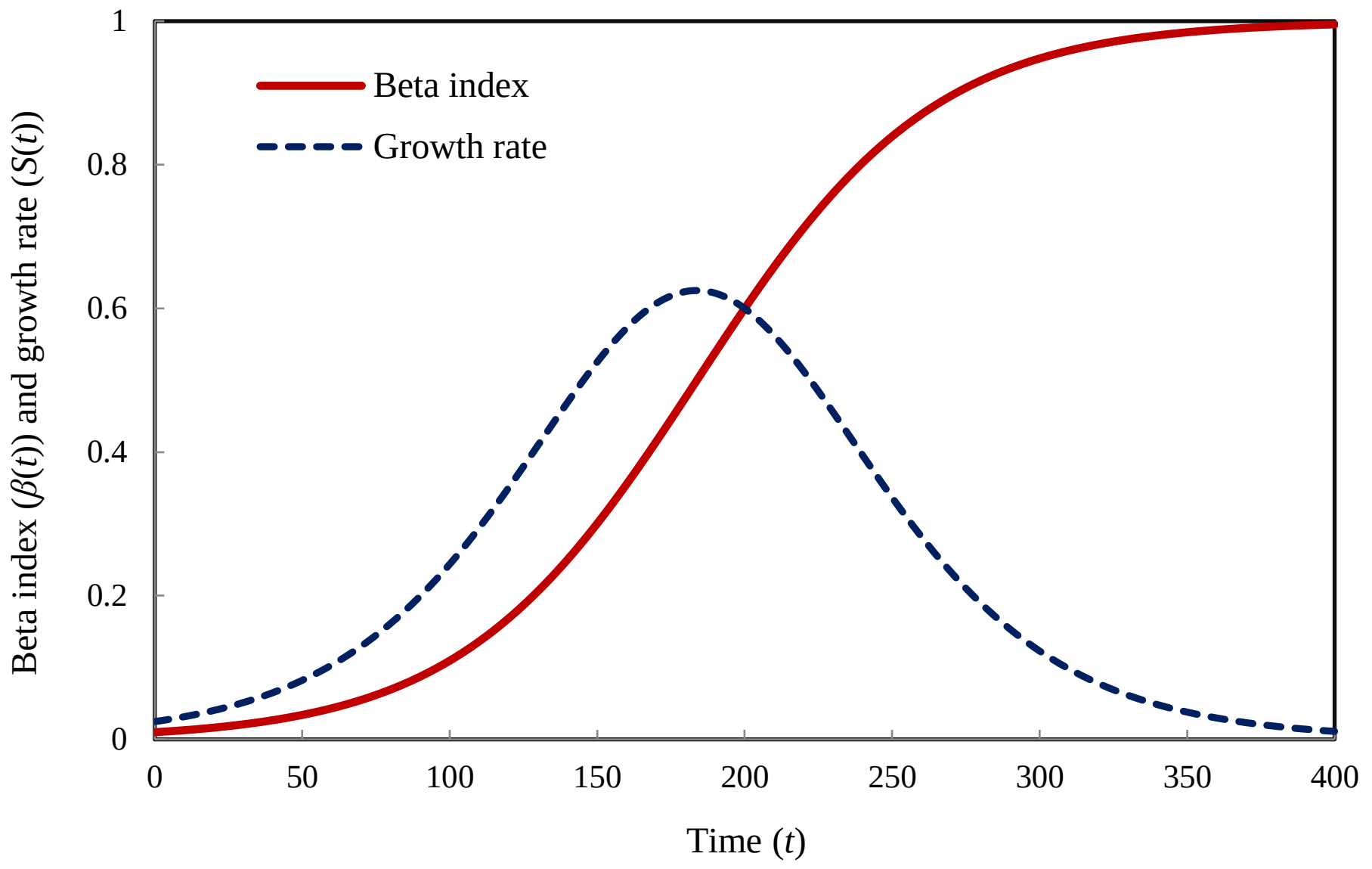

**Figure 3 Schematic diagram of the curves of $\beta$ index and its growth rate of traffic networks**

(**Note**: This is the curve based on the pure theoretical models. For the sake of intuition, the growth rate value is "magnified" by 100 times, otherwise the growth rate and the $\beta$ index cannot be displayed in the same coordinate diagram.)

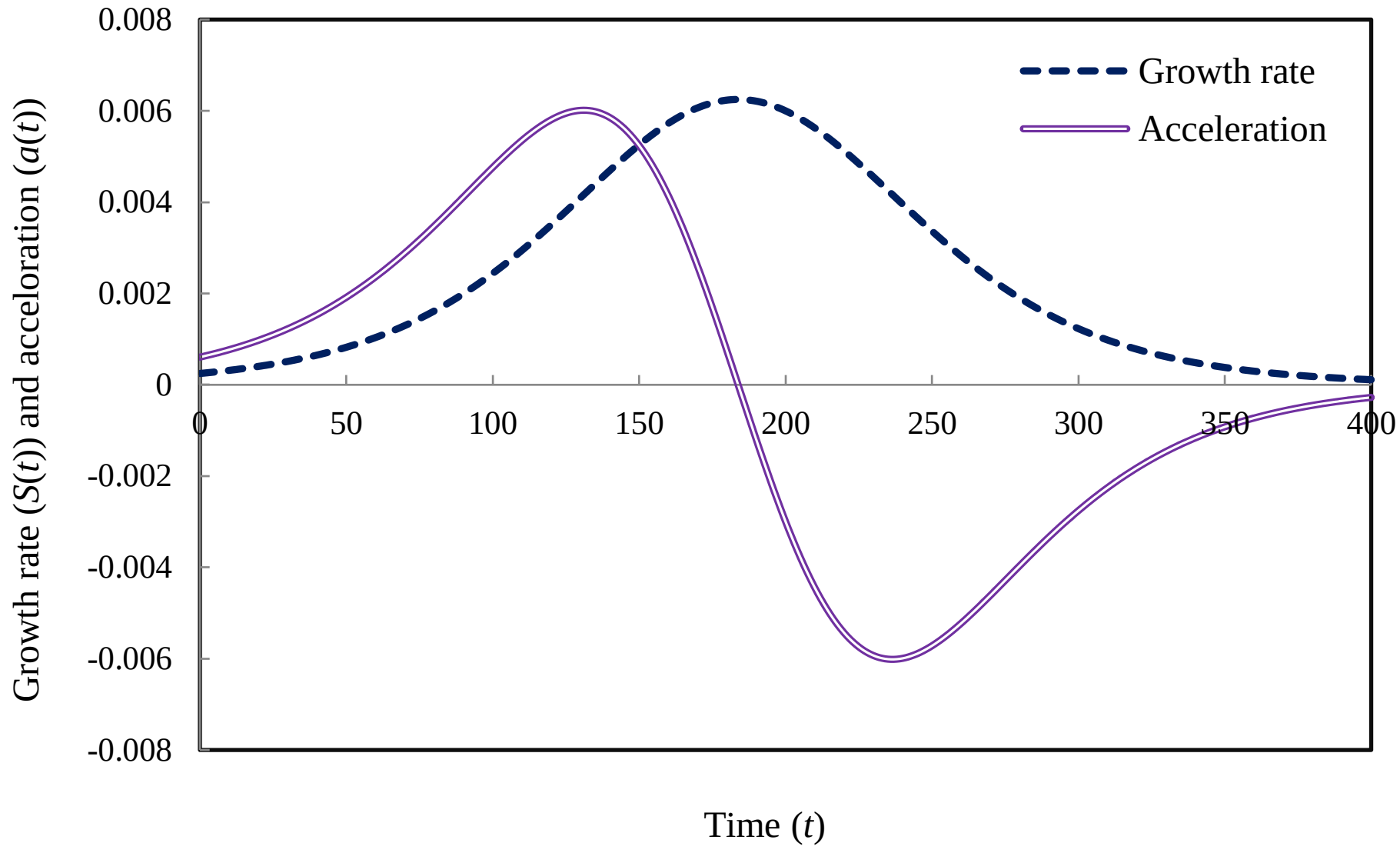

**Figure 4 Schematic diagram of the growth rate and acceleration curves of $\beta$ index of traffic network**

(**Note**: This is also the curve based on the pure theoretical models. For the sake of intuition, the acceleration value is "magnified" 100 times.)

The phase division model of the $\beta$ index growth can be employed to research mature traffic network, but not suitable for young traffic network. For the young traffic network, we can make use of the $\delta$ index to define the growth stage. Similar to equations (35) and (36), the growing speed and acceleration formulae for the $\delta$ index can be derived from equation (15) as below

$$S^{*}(t)=\frac{\mathrm{d}\delta(t)}{\mathrm{d}t}=k\delta(t)(1-\frac{\delta(t)}{\delta_{\max}}), \tag{37}$$

$$a^{*}(t)=\frac{\mathrm{d}^{2}\delta(t)}{\mathrm{d}t^{2}}=\frac{\mathrm{d}S^{*}(t)}{\mathrm{d}t}=k^{2}\delta(t)(1-\frac{2\delta(t)}{\delta_{\max}})(1-\frac{\delta(t)}{\delta_{\max}}), \tag{38}$$

where $S^{*}(t)$ and $a^{*}(t)$ denotes the growth rate and acceleration of the $\delta$ index at time $t$, respectively. Therefore, the growing process of a traffic network can also be divided into four stages based on the $\delta$ index: initial slow growth, accelerated fast growth, decelerated fast growth, and terminal slow growth (Table 2). The stage division patterns based on $\delta$ index is similar to that based on $\beta$ index, but the growth curve of $\beta$ index is not synchronized with the growth curve of $\delta$ index. The peak and valley values of the former lag behind those of the latter.

**Table 2 Phase division results and inflexion points of transport network development based on logistic model of the $\delta$ index**

| Scheme (I) | Scheme (II) | Scheme (III) | Middle limit | Lower and upper limits |
|---|---|---|---|---|
| Acceleration | Initial slow growth | Initial slow growth | | $\delta_{l}(t)=(\frac{1}{2}-\frac{1}{2\sqrt{3}})\delta_{\max}$ |
| | Celerity (fast growth) | Accelerated fast growth | $\delta_{m}(t)=\frac{\delta_{\max}}{2}$ | |
| Deceleration | | Decelerated fast growth | | $\delta_{u}(t)=(\frac{1}{2}+\frac{1}{2\sqrt{3}})\delta_{\max}$ |
| | Terminal slow growth | Terminal slow growth | | |

**Note**: In the table, $\delta_{l}(t)$ refers to the lower division, $\beta_{m}(t)$ to the middle division, and $\delta_{u}(t)$ to the upper division.

By means of the empirical models, we can generate two patterns of stage division of traffic development. Using equation (32), we can give a growth pattern of stage division of traffic network based on $\beta$ index (Figure 5). Using equation (33), we can generate a growth pattern of stage division of traffic network based on $\delta$ index (Figure 6). For each stage division scheme, the time span of the three inflection points of acceleration peak, growth rate peak, and acceleration valley is about 52 years. For the two stage division schemes, there is time a difference of about 39 to 40 years between corresponding peak values or valley values. For example, the peak of growth rate of *β* index lags behind the peak of growth rate of *δ* index by about 40 years. Studying these growth stages can provide many useful urban and traffic development information, but this is not the subject of this work.

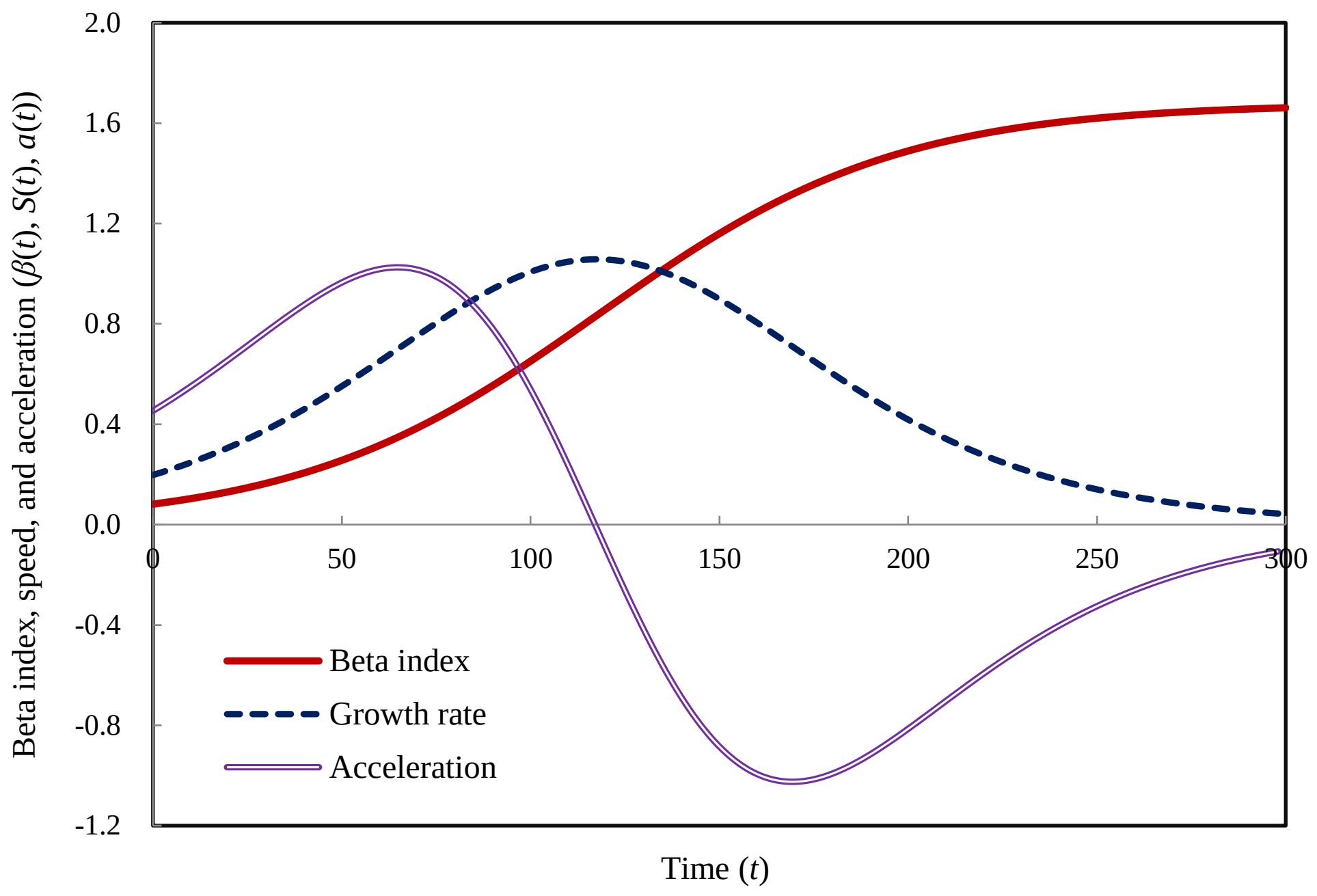

**Figure 5 A pattern of growth stage division based on the *β* index of traffic connectivity**

(**Note**: This is the family of curves based on equation (32). For the sake of intuition, the growth rate and acceleration values are "magnified" by 100 times.)

The theories and methods similar to this research have not been reported in the literature. There are three shortcomings in this study. First, the main result of theoretical derivation is two-parameter logistic function rather than three-parameter logistic function. The next step is to construct a dynamic equation for derivation of three-parameter logistic function. Of course, if we derive a logistic model based on normalized variable, the problem is easily to solve. Second, the theoretical and empirical analyses were limited to common logistic function and Boltzmann equation. In fact, the urbanization curves of European and American countries can be modeled with the conventional logistic function and Boltzmann equation, while Chinese and India urbanization curves should be modeled with quadratic logistic function and quadratic Boltzmann equation. In China, the urbanization model in the south is also different from that in the north. The urbanization curves in many regions of the South satisfy the conventional logistic function, while the urbanization curves in most regions of the North satisfy the quadratic logistic curve. If an urbanization curve satisfy a quadratic function or Boltzmann equation, the corresponding $\beta$ and $\delta$ index curves will also satisfy quadratic function or Boltzmann equation. Third, no direct and systemic empirical verification for the results of theoretical derivation due to absence of time series of observational data. It is easy to obtain a cross-sectional dataset for the $\beta$ index of traffic network, but is hard to get a long sample path of time series for this index. The next step is to carry out an empirical studies based on time series of the $\beta$ index of traffic network. The $\beta$ index can be transformed into $\delta$ index for prediction analysis and growth stage research on traffic network.

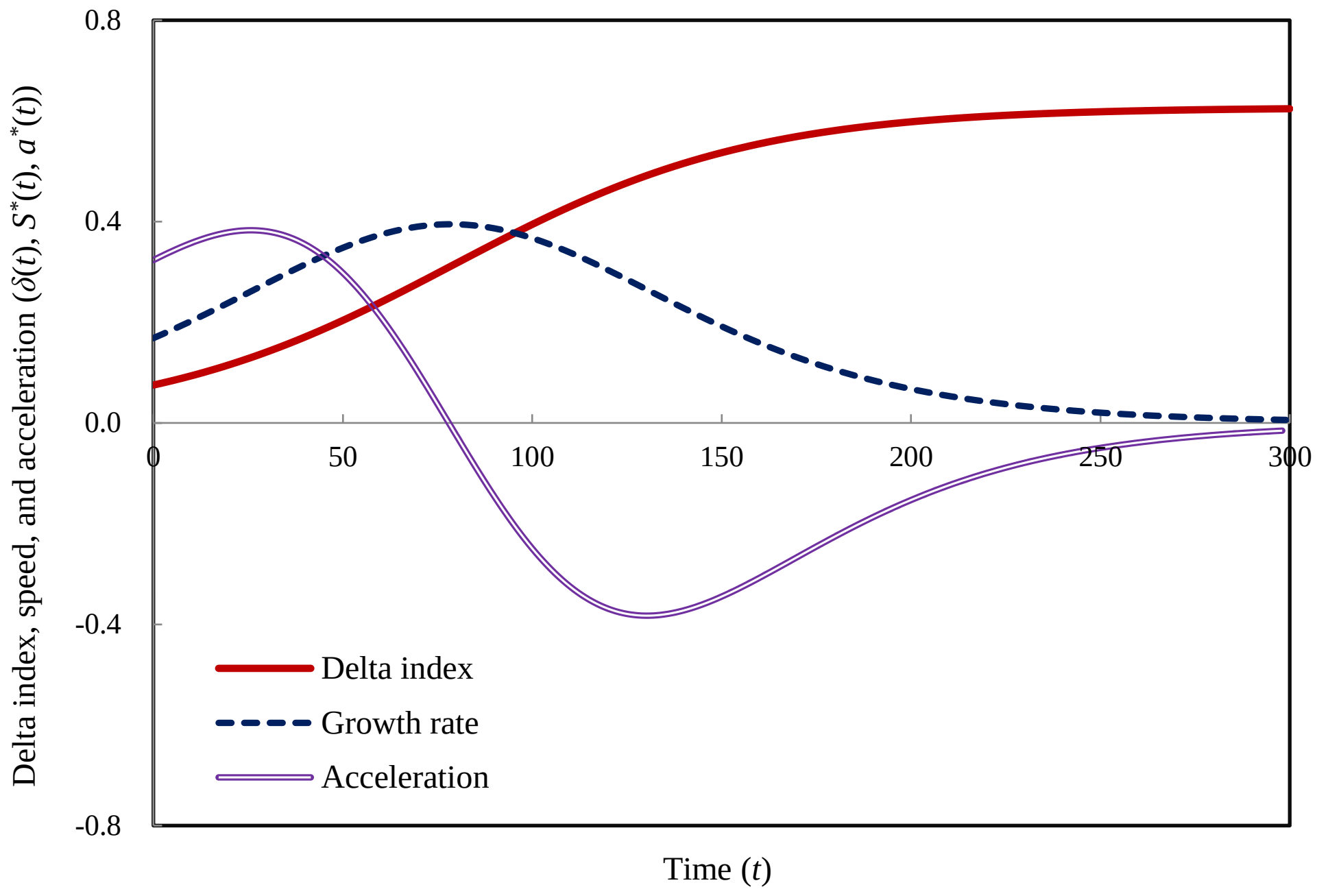

**Figure 6 A pattern of growth stage division based on the $\delta$ index of traffic connectivity**

(**Note**: This is the family of curves based on equation (33). For the sake of intuition, the growth rate and acceleration values are "magnified" by 100 times.)

# 5 Conclusions

This is a research achievement focusing on theoretical models and prediction methods. Positive analysis is not the chief aim of this work. The main conclusions can be drawn as follows. First, the growing curves of the $\beta$ index of traffic network in a geographical region can be described and predicted with a proper sigmoid function. The family of sigmoid functions include logistic function, generalized logistic function, hyperbolic tangent function, Boltzmann equation, quadratic Boltzmann equation, and Gompertz function. Where traffic networks are concerned, the $\beta$ index curves can be modeled with Boltzmann equation or quadratic Boltzmann equation, and a Boltzmann equation can be transformed into, reduced to, or approximated as a logistic function. Second, the sigmoid function of the $\beta$ index of traffic network can be derived from a pair nonlinear dynamic equations. The dynamic equations describe the coupling or interaction between human settlements and roads. In an urban system, human settlements (cities and towns) can be abstracted as nodes (points), and roads (highways, railways, or waterways) can be abstracted as edges (lines). Exploring the nonlinear relations and spatial interaction between nodes and edges may disclose the inherent mechanism of spatio-temporal evolution of urban and traffic networks. Third, the main uses of the sigmoid functions for the $\beta$ index of traffic network lie in explanation, prediction, and stage division of traffic network development. By the interaction of roads and human settlements, we can explain the dynamic mechanism of traffic network evolution. By means of Boltzmann equation or logistic function, we can predict the values of the $\beta$ index of a traffic network at given time. By using the growth rate formula and acceleration formula of the $\beta$ index based on logistic model, we can divide a growing process of a traffic network into four stages: initial slow growth, accelerated fast growth, decelerated fast growth, and terminal slow growth. Fourth, another measurement termed $\delta$ index can be derived from the $\beta$ index, and the two indices complement each other in traffic spatial analysis. All the models suitable for the $\beta$ index are suitable for $\delta$ index and *vice versa*. Compared with the $\beta$ index, the $\delta$ index bears clear lower and upper limits and easily to be used to make analysis. Where stage division of traffic network growth is concerned, the $\delta$ index is suitable for young network, while the $\beta$ index is suitable for mature network. There is a time lag between the phase division curve based on the $\beta$ index and that based on the $\delta$ index. The values of time lag may show useful geographical information for analyzing spaito-temporal evolution of traffic networks.

## Acknowledgement

This research was sponsored by the National Natural Science Foundation of China (Grant No. 42171192). The support is gratefully acknowledged.

# Appendix

## A1 Derivation of equation (30)

A Boltzmann equation about traffic connectivity can be derived from the linear relation between the $\beta$ index and urbanization level as well as the logistic model of the $\beta$ index growth. Let

$$\omega L(t)=\beta(t)-\beta_{\min}, \tag{A1}$$

we have

$$\omega L_{\max}=\beta_{\max}-\beta_{\min},\quad \omega L_0=\beta_0-\beta_{\min}. \tag{A2}$$

Further, let $\vartheta=\beta_{\min}$, we have $\omega L_{\max}=\beta_{\max}-\beta_{\min}$, and accordingly $\omega L_0=\beta_0-\beta_{\min}$, then we have

$$L_{\max}=\frac{\beta_{\max}-\beta_{\min}}{\omega},\quad L_0=\frac{\beta_0-\beta_{\min}}{\omega}. \tag{A3}$$

Substituting $\vartheta=\beta_{\min}$ and equation (A3) into equation (28) yields

$$\begin{aligned}\beta&=\beta_{\min}+\frac{\beta_{\max}-\beta_{\min}}{1+((\beta_{\max}-\beta_{\min})/\omega/(\beta_0-\beta_{\min})/\omega-1)e^{-kt}}\\&=\beta_{\min}+\frac{\beta_{\max}-\beta_{\min}}{1+(\frac{\beta_{\max}-\beta_0}{\beta_0-\beta_{\min}})e^{-kt}}\end{aligned}, \tag{A4}$$

which is just the result that we need.

## A2 Derivation of equations (32) and (33)

Two empirical models of logistic growth for $\beta$ index and $\delta$ index are derived from the previous published empirical models in the text. The more detailed mathematical process is as follows. Substituting equation (26) into equation (20) yields

$$\hat{\beta} = 0.197 + 0.165\ln(0.5z^{1.05}) = 0.0823+0.1737\ln z\,. \tag{A5}$$

Equation (22) can be re-expressed as

$$\ln z = \frac{\hat{L} + 74.6799}{17.5694}\,. \tag{A6}$$

Substituting equation (A6) into equation (A5) yields

$$\hat{\beta} = 0.0823+0.1737\frac{(L + 74.6799)}{17.5697} = 0.8208+0.0099L\,. \tag{A7}$$

Substituting equation (24) into equation (A7) yields

$$\hat{\beta} = 0.8208+\frac{0.8572}{1+19.6016e^{-0.0252t}}\,. \tag{A8}$$

In light of Boltzmann model, we have

$$\hat{\beta} = 0.8208\,, \hat{\beta}_{\max} - \hat{\beta}_{\min} = 0.8572\,. \tag{A9}$$

Thus $\beta_{\max}$=0.8572+0.8208=1.6780. The Boltzmann model can be reduced to logistic model as below:

$$\hat{\beta}_{\min} \approx \frac{1.6780}{1+19.6016e^{-0.0252t}}\,. \tag{A10}$$

According to the mathematical structure of logistic models, we have

$$\hat{\beta}_{\max} = 1.6780\,,\ \frac{\hat{\beta}_{\max}}{\hat{\beta}_0} - 1 = 19.6016\,. \tag{A11}$$

So the initial value of the $\beta$ index is

$$\hat{\beta}_0 = \frac{\hat{\beta}_{\max}}{20.6016} = \frac{1.6780}{20.6016} = 0.0815\,. \tag{A12}$$

In terms of equations (5) and (6), the capacity value and initial value of the $\delta$ index are

$$\hat{\delta}_{\max} = \frac{\hat{\beta}_{\max}}{1+\hat{\beta}_{\max}} = \frac{1.6780}{2.6780} = 0.6266\,, \tag{A13}$$

$$\hat{\delta}_0 = \frac{\hat{\beta}_0}{1+\hat{\beta}_0} = \frac{0.0815}{1+0.0815} = 0.0753\,. \tag{A14}$$

Therefore, we have

$$\frac{\hat{\delta}_{\max}}{\hat{\delta}_0}-1=\frac{0.6266}{0.0753}-1=7.3194. \quad \text{(A15)}$$

Substituting $k$=0.0252 and equations (A13) and (A15) into equation (15) yields equation (33).

## A3 Logarithmic relations between urbanization and economic development levels

The following figures show two examples of the logarithmic relationship between the level of urbanization and that of economic development. The examples lend further support to the inference in Subsection 3.2. The original data for Figure A(a) come from World Development Indicators 2000, The World Bank, which includes the data of urbanization level and per capita GDP of 174 countries all over the world in 1998. Based on per capita GDP, the average values are calculated by taking 5 countries as a group. The last group comprises 4 countries. The original data for Figure A(b) come from World Development Indicators 2002, The World Bank, which includes the data of urbanization level and per capita GDP of 137 countries in 2000. The average values are calculated by taking 4 countries as a group. The last group comprises only 1 country.

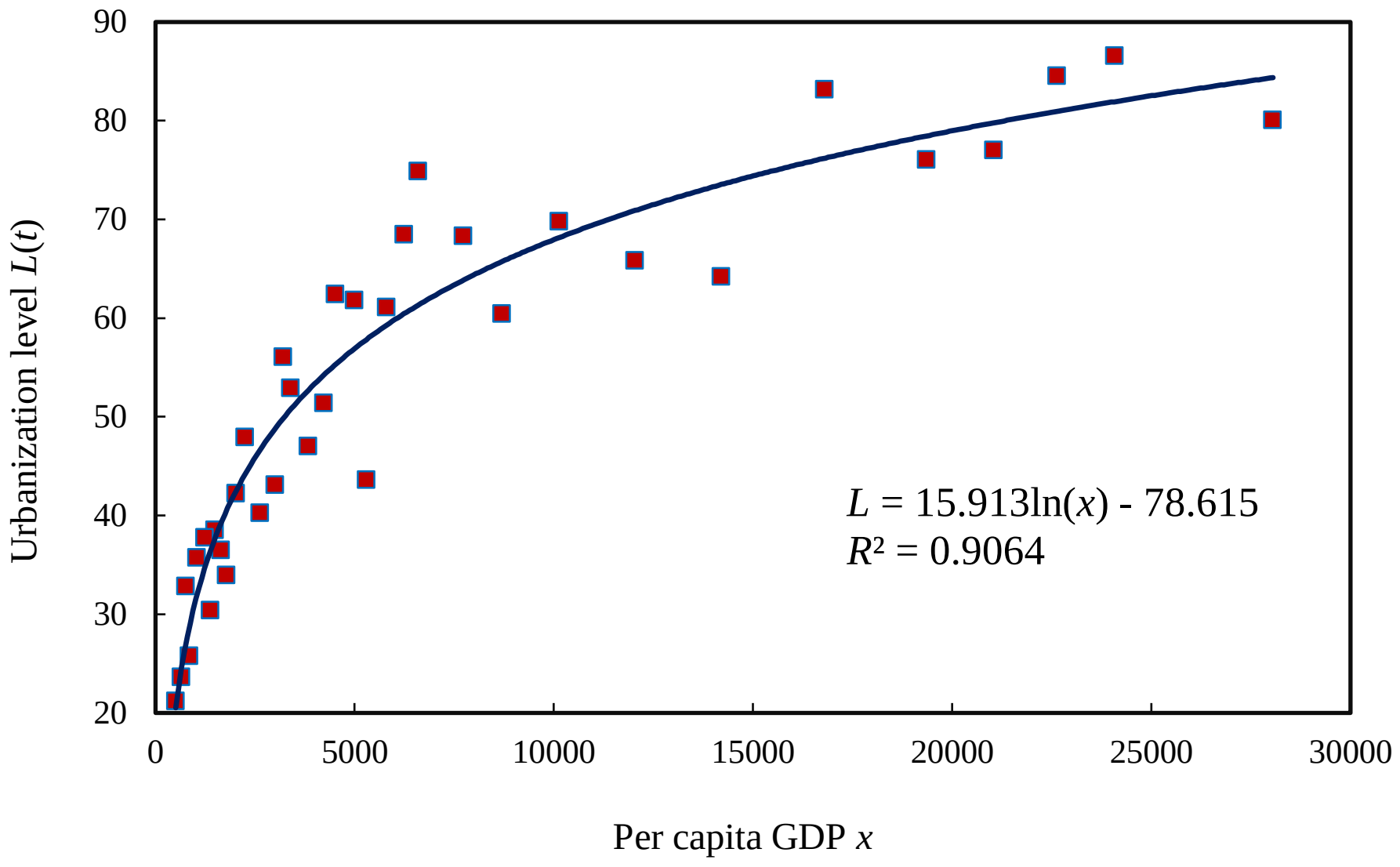

a. 174 countries in 1998

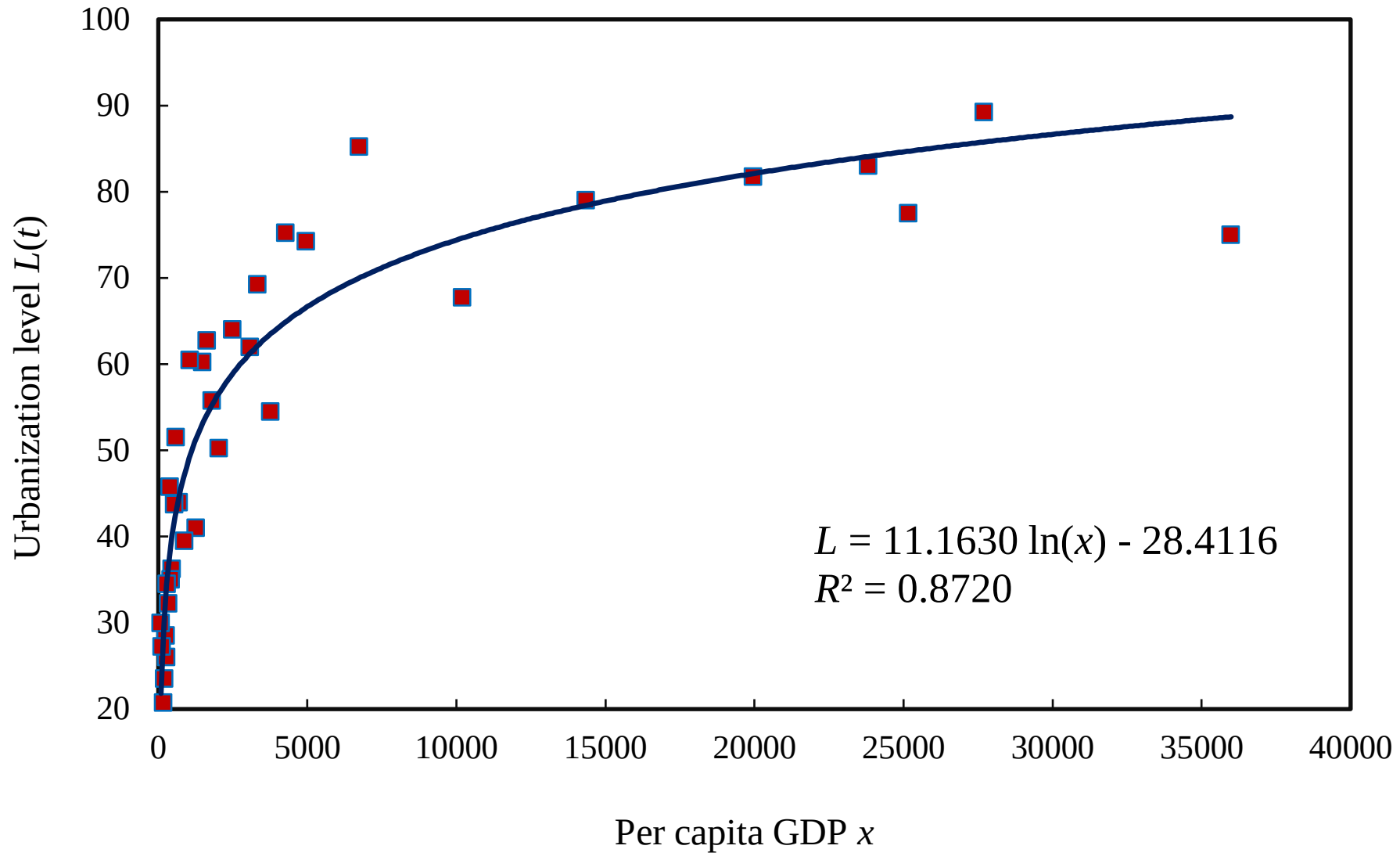

a. 137 countries in 2000

**Figure A The logarithmic relationship between urbanization level and economic development level based on per capita GDP of the countries around the world**